# A Brief AI Literacy Intervention Does Not Significantly Reduce Over-Reliance and Increases Under-Reliance on ChatGPT: A Randomized Study

Brett Puppart, Jaan Aru
Institute of Computer Science, University of Tartu
brett.puppart@gmail.com

## Abstract

In this study, we examined whether a brief AI literacy intervention influences high school students' reliance on recommendations from large language models (LLMs). In a randomized experiment, students were assigned to either a control group receiving a brief introduction to LLMs or an intervention group receiving additional information about how LLMs work, their limitations, and effective usage strategies. Participants then solved eight math puzzles with ChatGPT's advice, which was incorrect in half of the trials. Results indicated widespread over-reliance, with incorrect recommendations adopted in 52.1% of the trials. The intervention did not significantly reduce over-reliance. Instead, it led to an increase in under-reliance, as students were more likely to reject correct recommendations. These findings provide preliminary evidence that brief text-based interventions may be ineffective in fostering appropriate reliance. More comprehensive and interactive approaches may be required to meaningfully influence students' real-world reliance on LLMs.

# 1. Introduction

Large language models (LLMs) are probabilistic models that have been trained to predict the next token in a text sequence (Naveed et al., 2024). High capabilities, ease of use, and commercial promotion have caused a widespread adoption of these systems in various fields, such as medicine (Thirunavukarasu et al., 2023), science (Liang et al., 2024), and increasingly education (Valova et al., 2024).

Researchers have suggested that LLMs could revolutionize education by enabling personalized learning, providing immediate feedback, and improving accessibility (Diab Idris et al., 2024; Wang et al., 2024). Reflecting this trend, Estonia launched the national "AI Leap" initiative in 2025 to introduce LLM-based tutors to high school students (ERR, 2025; Aru & Laak, 2025). Similar initiatives have emerged in other countries, signaling the growing integration of LLMs into educational systems (OpenAI, 2026).

Despite these potential benefits, LLM use also introduces risks. One substantial risk is over-reliance, in which people adopt incorrect AI recommendations (Buçinca et al., 2021). Many humans display a general aversion towards cognitive effort (Aru & Rozgonjuk, 2022; David et al., 2024; Shenhav et al., 2017). In our interaction with external tools, this tendency may increase cognitive offloading, defined as taking physical action to reduce task demand (Risko & Gilbert, 2016). Students may be tempted to delegate parts of thinking to LLMs, which can provide immediate solutions to many problems. Consistent with this expectation, several studies have reported over-reliance in educational contexts (Amoozadeh et al., 2024; Valova et al., 2024).

Such over-reliance might contribute to several problems at both societal and individual levels. At the societal level, over-reliance may contribute to the spread of harmful misinformation and biases. LLMs often present answers in a very confident style (Azaria et al., 2024), which is problematic as they may contain errors (Borji, 2023) and produce biased or stereotypical outputs (Kotek et al., 2023; Rozado, 2023). At the individual level, reliance on LLMs may reduce engagement in deliberate reasoning and critical thinking (Aru, 2025; Dubey et al., 2024; Kasneci et al., 2023). Over time, such external dependency might harm our critical thinking (Duhaylungsod & Chavez, 2023; Zhai et al., 2024; Gerlich, 2025), metacognitive regulation (Fan et al., 2025; Fernandes et al., 2024; Laak et al., 2024), and information retention (Sparrow et al., 2011; Dubey et al., 2024).

AI literacy interventions have been proposed as a potential way to mitigate over-reliance on AI systems (Kasneci et al., 2023; Morrill & Noetel, 2023). Broadly defined, AI literacy refers to the competencies required to critically evaluate, collaborate with, and use AI in different areas of life (Long & Magerko, 2020). While some AI literacy frameworks include affective and enacted components, such as attitudes and self-reflection (Ng et al., 2021; Miao & Shiohira, 2024; OECD, 2025), we follow the line of work that conceptualizes AI literacy as knowledge and skills (Chiu et al., 2024; Chiu, 2025).

Despite interest in the topic, the relationship between AI literacy and over-reliance in LLM-human collaboration is not well understood. Previous research in broader machine-learning and automation contexts suggests that greater understanding of AI can influence, and in some contexts reduce, over-reliance (Chiang & Yin, 2021; Horowitz & Kahn, 2023; Jacobs et al., 2021; Morrill & Noetel, 2023). However, preliminary work in the LLM context suggests that the association may be more complex. Zhang et al. (2025) found that greater AI literacy was associated with greater AI dependency among preservice mathematics teachers. AI dependency, in turn, was negatively associated with several 21st-century skills, including critical thinking and problem-solving. These findings suggest that greater knowledge or proficiency with AI might not translate into more cautious or appropriate reliance on LLMs.

Even if AI literacy promotes more appropriate reliance, the amount of training or instruction required to induce measurable effects remains unclear. Schools adopting AI tools may favor brief informational interventions, such as short explanatory texts or brochures, over longer courses or more extensive materials because they are easier to implement at scale. Some experimental work suggests that brief interventions may reduce over-reliance. Chiang and Yin (2021) found that informing participants about possible variation in model performance reduced reliance during covariate shift, while Morrill and Noetel (2023) found that a short AI literacy intervention reduced over-reliance on AI advice. More recently, Bo et al. (2024) found that several interventions reduced over-reliance on LLM advice, although they generally failed to promote appropriate reliance. However, little evidence exists on whether brief AI literacy interventions can improve high school students' reliance on LLM advice.

This study aims to fill this gap by presenting preliminary experimental findings on the effectiveness of a brief LLM-specific intervention on suboptimal reliance. High school students were randomly assigned to either a control or intervention group and completed a behavioral task involving acceptance or rejection of ChatGPT recommendations. We address the following research questions:

**RQ1**: How does a brief AI literacy intervention impact high school students' acceptance of incorrect ChatGPT recommendations (over-reliance)?

**RQ2**: How does a brief AI literacy intervention impact high school students' acceptance of correct ChatGPT recommendations (under-reliance)?

Additionally, given prior work linking over-reliance to reduced deliberation (Buçinca et al., 2021), we examined whether decision time was associated with reliance behavior. We also explored whether the intervention influenced participants' metacognitive confidence in their responses.

## 2. Method

### 2.1. Participants

The sample comprised 12th-grade Estonian high school students (*M* = 18.06 years, *SD* = 0.23, age range: 18−19 years). The invitations to participate in the study were sent to 72 schools, of which 10 participated in the study and forwarded the study invitation to their senior students. A total of 42 students completed the questionnaire. Six of them were excluded from the analysis for the following reasons: Five participants were excluded because they read the intervention text in less than 30 seconds, and one participant was excluded because they completed the entire questionnaire in less than six minutes. Therefore, the analytic sample consisted of 36 participants, of whom 25 (69.4%) were female and 11 were male (30.6%). To incentivize participation, a lottery was held among all participants, with five €20 bookstore gift cards as prizes.

### 2.2. Materials

For the intervention group, we created a brief educational text (hereafter, *intervention text*). The content was based on Walter's (2024) conceptualization of AI literacy, given its emphasis on the limitations and risks of AI. The intervention text covered all sub-skills of AI literacy, except for ethics, as this aspect is addressed in separate guidelines (Table 1). Screenshots illustrating ChatGPT's hallucinations and other limitations were also provided. For the control group, we created a text (hereafter, *control text*) that contained only a general introduction to ChatGPT, which was also presented in the intervention group (Table 1). Full translations of the texts from Estonian are provided in Appendix B.

| Section | Content description | Control | Intervention |
|---|---|---|---|
| Introduction | Overview of ChatGPT as a text-based system and its capabilities | Included | Included |
| Working mechanism | Explanation of LLMs as probabilistic models predicting the next token | Not included | Included |
| Limitations and risks | Description of hallucinations, inaccuracies, and potential biases | Not included | Included |
| Prompting and usage strategies | Guidance on effective use of ChatGPT and prompt formulation | Not included | Included |

**Table 1.** *Overview of the texts presented in the control and intervention groups.*

To measure reliance behavior, we used Kangaroo math puzzles (Teaduskool, 2022), which assessed not only the knowledge of formulas but also logical and creative reasoning. These puzzles were suitable for measuring over-reliance for two main reasons. First, each puzzle has one correct answer, allowing over-reliance to be classified unambiguously as a binary outcome. Second, solving these puzzles requires iterative reasoning and slow, analytical thinking. This aspect mimics the real-life use of ChatGPT, where fact-checking requires additional effort compared to uncritical acceptance.

To generate suggestions, we used ChatGPT-4o with default parameters. We did not explicitly prompt the model to err. Instead, we added the puzzle to the query with the instruction "solve correctly". This ensured that the incorrect recommendations arose naturally from the model's responses, thereby increasing ecological validity.

We tested several puzzles from the "Cadet" age category, designed for 7th-8th grade students, to ensure a consistent and appropriate level of difficulty for 12th-graders. We selected eight puzzles, with four featuring correct recommendations from ChatGPT and four featuring incorrect recommendations (Appendix A). Simple puzzles were avoided to encourage engagement with ChatGPT's recommendations. Incorrect recommendations had to appear trustworthy but contain errors that participants should be able to detect through critical evaluation.

## 2.3. Procedure

The study was conducted on the LimeSurvey online platform, with data collection taking place in October 2024. Upon entering the study, participants provided informed consent and basic demographic information, including gender, age, and school affiliation.

Participants were then randomly assigned to either the intervention or the control group and received the corresponding text. Following the reading task, participants completed a brief manipulation check, assessing how well they perceived the text had informed them about three aspects of ChatGPT: how it works, its limitations and risks, and how to use it effectively. Responses were collected on a four-point Likert scale.

After that, participants received instructions and proceeded to solve math puzzles with ChatGPT's assistance. The instructions were as follows: "You are about to solve math puzzles with the help of ChatGPT. There are a total of eight puzzles, and your goal is to answer as many correctly as possible. If you choose not to accept the recommendation, that option will not be available later. There is no time limit."

Each participant was presented with eight puzzles in random order. For each puzzle, participants were first shown a screenshot of ChatGPT's recommendation (Figure 1). Participants indicated whether they would use it as their answer. If they selected "yes," they proceeded to the next puzzle. If they selected "no," they solved the puzzle independently by selecting an answer from four options (ChatGPT's recommendation was removed). After each puzzle, participants rated their confidence on a five-point Likert scale (1 = strongly disagree – 5 = strongly agree) in response to the statement, "I believe my answer to the previous puzzle was correct."

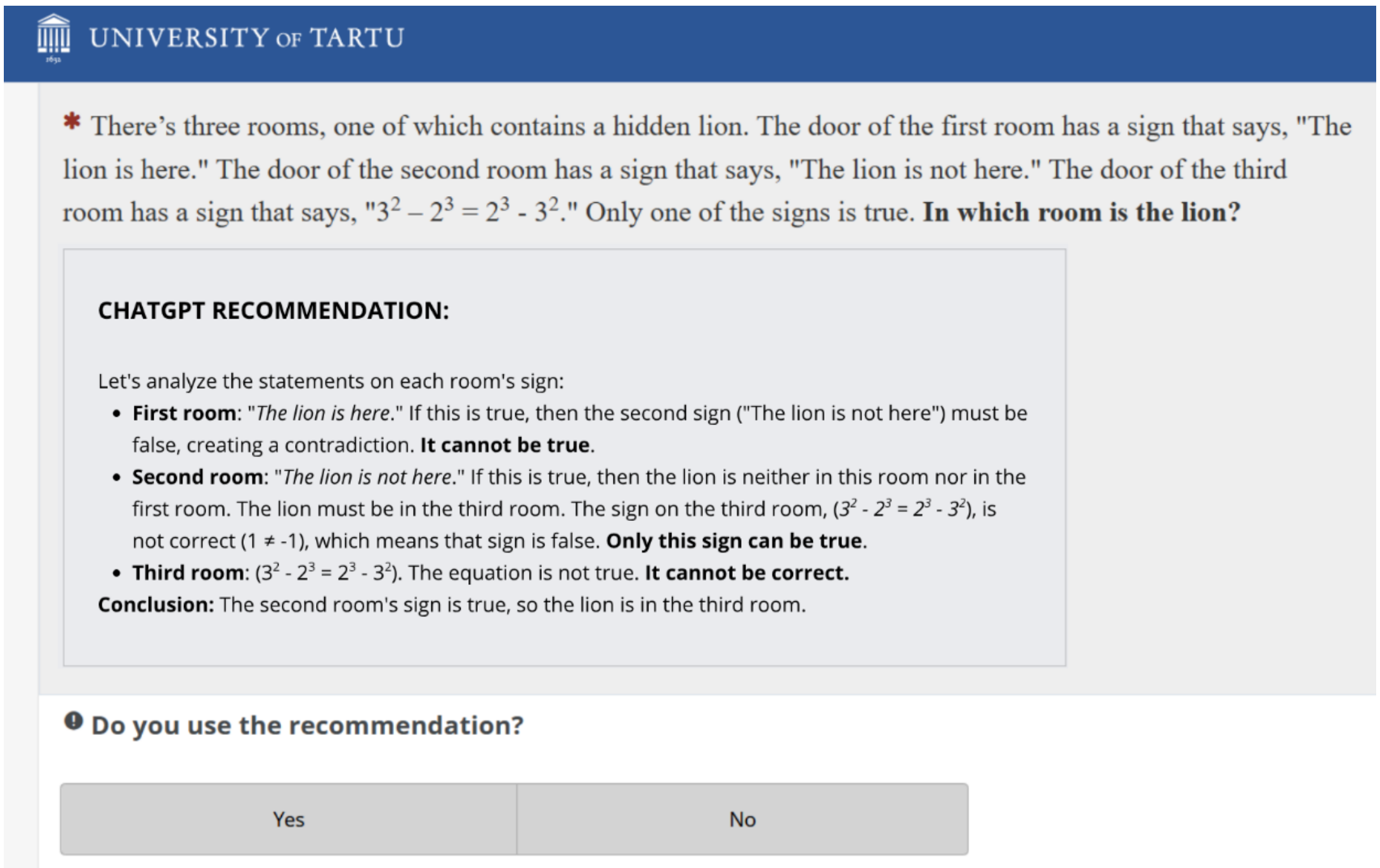

**Figure 1.** *Example of a puzzle presented to the participants*

The primary dependent variable of the study was over-reliance, which we operationally defined as a scenario where ChatGPT provided an incorrect recommendation, and the participant adopted it as their answer. We also measured under-reliance, which we operationally defined as a scenario where the participant did not adopt ChatGPT's correct recommendation.

Additionally, we measured the time taken to decide whether to rely on ChatGPT's recommendation. We used three different metrics for these purposes: decision time across all puzzles; decision time for puzzles with incorrect recommendations; and decision time for puzzles with correct recommendations. Beyond decision time, we also measured the total time spent reading the texts and the total time spent solving the puzzles. All time metrics were reported in minutes.

## 2.4. Ethics

This study was approved by the Research Ethics Committee for Human Studies at the University of Tartu. All participants read detailed instructions and provided informed consent before taking part in the study. The study was presented to participants under the title "Students' motivation in solving mathematics puzzles with the help of ChatGPT." This was done to prevent bias – if participants were aware of the actual aim of the study, it might have compromised the validity of the results. At the end of the experiment, a debriefing was conducted in which participants were informed about the actual purpose of the study. This ensured the ethical integrity of the research.

## 2.5. Data analysis

The data analysis was conducted using the open-source statistical software JASP (version 0.19.1) to examine the relationships between variables. The significance level for all analyses was set at $p = .05$. Normality assumption was considered violated when the absolute value of skewness or kurtosis exceeded 2, and homogeneity of variance was considered violated when Levene's test was statistically significant.

To compare reliance, decision time, confidence, and manipulation effectiveness between groups, we applied the independent samples t-test. When normality was violated, we applied the non-parametric Mann-Whitney U test. When homogeneity of variance was considered violated, we used Welch's t-test. Cohen's $d$ was used as the primary effect size for independent samples and Welch's t-test, while for Mann-Whitney U test, rank-biserial correlation was used ($r_{rb}$). We interpreted effect sizes according to Cohen's (1988) guidelines: for Cohen's *d*, 0.20 represents a small effect, 0.50 a medium effect, and 0.80 a large effect. For $r_{rb}$ we interpreted the results according to standard correlation guidelines, described below.

To analyze the relationship between decision time and reliance, we conducted correlation analyses. In case of normality violation, we used Spearman's rank correlation coefficient (ρ). Effect sizes for $r$, $\rho$ and $r_{rb}$ were interpreted using Cohen's (1988) guidelines, where values of .10, .30, .50 represent small, medium, and large effects, respectively.

## 3. Results

The analytic sample consisted of 36 students, with 19 in the intervention group and 17 in the control group. Participants completed the study in 23.6 minutes on average; the intervention text required more reading time (4.3 min) than the control text (0.9 min).

Over-reliance occurred in 52.1% ($SD$ = 28.90) of trials with incorrect recommendations, and under-reliance in 28.5% ($SD$ = 24.00) of trials with correct recommendations. Most participants accepted at least one incorrect recommendation (94.4%) and rejected at least one correct recommendation (75.0%). Appendix C shows that over-reliance was similar across tasks, whereas under-reliance varied more substantially.

Welch's t-test indicated that over-reliance was lower in the intervention group ($M$ = 48.68, $SD$ = 24.26) than in the control group ($M$ = 55.88, $SD$ = 33.69), but the difference was not statistically significant, $t(28.78) = -0.73$, $p = .471$, $d = -0.25$ (Figure 2). An independent samples t-test indicated that under-reliance was higher in the intervention group ($M$ = 36.84, $SD$ = 24.11) than in the control group ($M$ = 19.12, $SD$ = 20.78), and this difference was statistically significant, $t(34) = 2.35$, $p = .025$, $d = 0.78$.

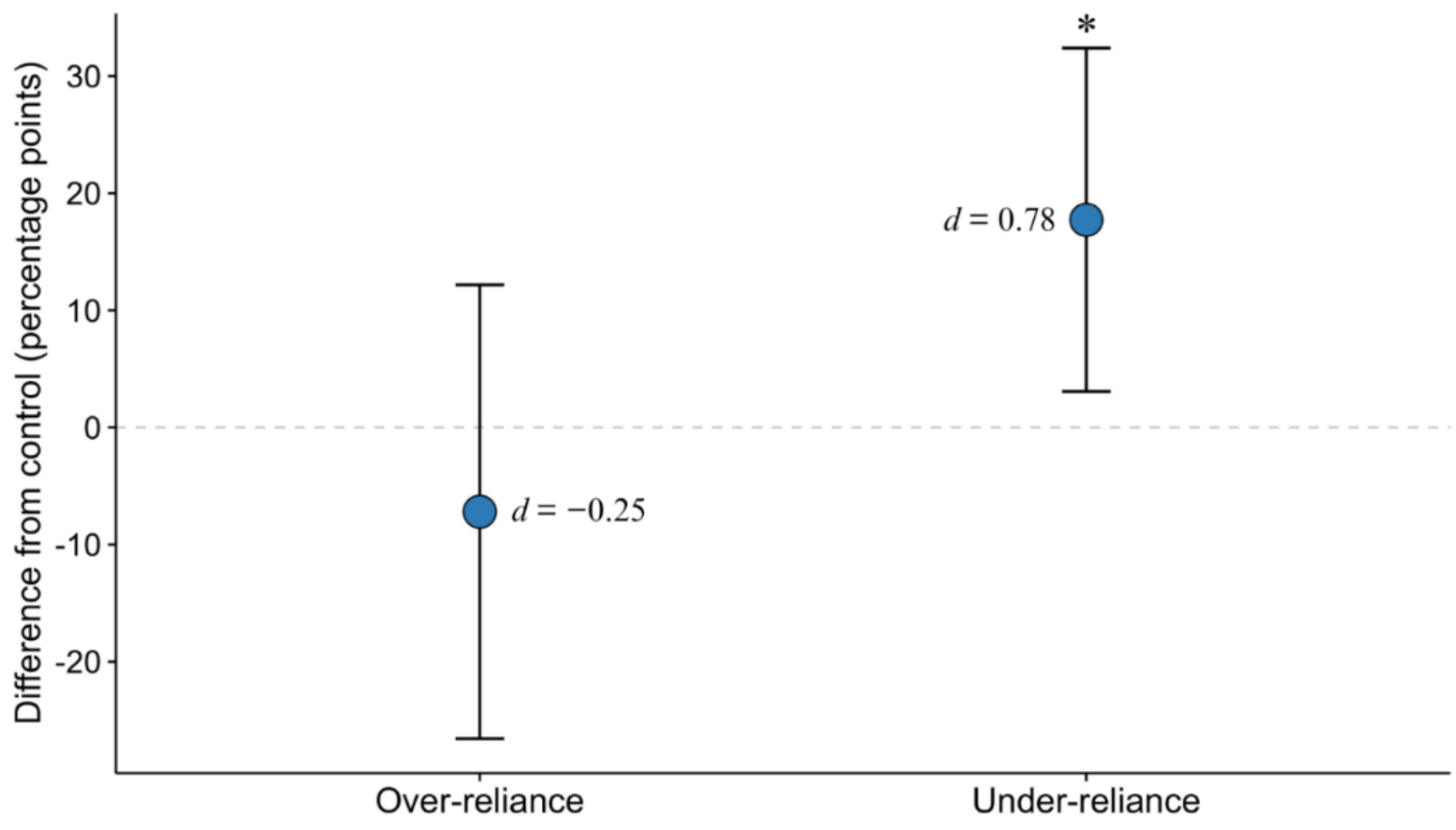

**Figure 2.** *Differences in over- and under-reliance between the intervention and control groups (n = 36)*. Values represent the mean difference (intervention − control) in percentage points. Over-reliance reflects the proportion of incorrect recommendations that were adopted, whereas under-reliance reflects the proportion of correct recommendations that were rejected. Error bars indicate 95% confidence intervals. The dashed horizontal line denotes no difference between groups. Asterisk (*) indicates a statistically significant difference ($p < .05$). Group differences for over-reliance were tested using Welch's t-test and for under-reliance using an independent samples t-test.

Independent samples t-tests indicated that participants in the intervention group reported higher perceived knowledge in the "Usage" category ($M = 3.42$, $SD = 0.77$) than in the control group ($M = 2.77$, $SD = 0.97$), $t(34) = 2.26$, $p = .030$, $d = 0.76$. Similarly, perceived knowledge in the "Working mechanism" category was higher in the intervention group ($M = 3.37$, $SD = 0.76$) than in the control group ($M = 2.77$, $SD = 0.90$), $t(34) = 2.09$, $p = .044$, $d = 0.70$. The difference in the "Limitations" category was not statistically significant, $t(34) = 0.56$, $p = .577$, $d = 0.19$. No significant group differences were observed in confidence ratings or decision time.

Over-reliance was negatively associated with decision time on trials with incorrect recommendations ($\rho = -0.41$, $p = .013$), whereas under-reliance was positively associated with decision time on trials with correct recommendations ($\rho = 0.35$, $p = .038$).

## 4. Discussion

The primary aim of this study was to assess the impact of a brief AI literacy intervention on high school students' reliance on recommendations provided by ChatGPT.

Results demonstrated widespread over-reliance, as students accepted nearly half of the incorrect ChatGPT recommendations. This is consistent with previous research suggesting that students often do not verify the accuracy of ChatGPT's outputs and place substantial trust in them (Amoozadeh et al., 2024; Valova et al., 2024). We found a negative association between over-reliance and decision times when evaluating incorrect ChatGPT recommendations. This is consistent with prior work showing that cognitive forcing functions can reduce over-reliance (Buçinca et al., 2021), and with the possibility that rapid, relatively automatic responding contributes to over-reliance (see Kahneman, 2011).

Considered alongside research into the consequences of external dependence on AI, our findings highlight the importance of designing and critically evaluating potential interventions against over-reliance. Previous work suggests that LLM use may reduce critical thinking (Gerlich, 2025), cognitive engagement (Kosmyna et al., 2025), learning (Bastani et al., 2025), and metacognitive calibration (Fernandes et al., 2024). While adoption of incorrect recommendations does not directly imply that any of these capabilities were harmed, it serves as an indirect behavioral proxy for how much students engaged with AI's feedback. Humans are naturally inclined to avoid cognitive effort (Aru & Rozgonjuk, 2022; David et al., 2024; Shenhav et al., 2017). When AI is routinely used without reflection, these cognitive capabilities may therefore suffer.

The key finding of this study is that brief AI literacy interventions may not be effective in reducing over-reliance on LLMs. Over-reliance was slightly lower in the intervention group, but this difference was not statistically significant. At the same time, the intervention significantly increased under-reliance. This pattern is consistent with the broader observation that interventions reducing over-reliance do not necessarily improve appropriate reliance overall (Bo et al., 2024). Our findings contrast with earlier work showing that brief informational interventions can reduce over-reliance on AI advice (Chiang & Yin, 2021; Morrill & Noetel, 2023).

One possible explanation for why our intervention increased under-reliance while showing no measurable effect on over-reliance is that it induced a general aversion towards AI advice, without improving participants' ability to selectively distinguish correct from incorrect recommendations. Previous research has shown that exposure to AI flaws can cause a general aversion towards its outputs, even when they are correct (Dietvorst et al., 2015). Consistent with this interpretation, there were no significant differences between groups in reliance decision time, suggesting that the intervention may not have increased deliberation. Instead, participants may have become generally more likely to dismiss ChatGPT's recommendations.

If replicated in larger studies, these findings would have important implications for educators. Briefly providing information about LLMs may be insufficient to promote appropriate reliance, even when students are explicitly informed about errors and biases. More extensive AI literacy programs may therefore be needed to meaningfully change students' reliance on LLMs.

## 5. Limitations

Several limitations should be considered regarding the methodology of this study. First, the actual change in AI literacy was not directly measured. As a result, it remains unclear to what degree students' knowledge about LLMs was influenced by the intervention. Although students in the intervention group reported increased knowledge in two out of three aspects compared to the control group, this relied on self-report, and no pre-post measure was included. Consequently, we cannot determine how much the intervention changed actual AI knowledge or skills, or whether participants already possessed much of the information presented. Future studies would therefore benefit from including pre- and post-intervention measures of AI literacy, ideally performance-based assessments, to determine whether the included interventions meaningfully increase AI literacy.

Another limitation concerns the ecological validity of the study environment. The artificial workflow of first introducing participants to a problem and then to the AI's solution may not adequately capture how LLMs are used in real-world contexts. The design might have inadvertently invited more metacognitive reflection by prompting participants to choose whether to use AI's advice, whereas users are not typically asked to make an explicit reliance decision after every recommendation in everyday use. The workflow also prevented participants from prompting, clarifying, or iteratively interacting with the model as they normally might.

The reliance task was also restricted to eight mathematics and logic puzzles with objectively correct answers. Reliance behavior may differ in more open-ended educational tasks, in domains where students have less prior knowledge, or when the quality of AI outputs is more difficult to verify. The fixed proportion of correct and incorrect recommendations may also differ from students' everyday experience with LLMs. Future studies should therefore test whether these findings generalize across task domains and interaction settings.

Finally, the sample size was relatively small, which may have limited statistical power to detect smaller effects. Future research should replicate these findings with larger samples to more reliably assess the effectiveness of brief AI literacy interventions.

## 6. Conclusion

In conclusion, identifying effective interventions for supporting appropriate reliance on LLMs is essential, as these tools are increasingly integrated into educational settings. One proposed solution is to increase students' AI literacy. Due to the ease of implementation and scalability, schools might be inclined to use brief informational approaches such as short texts or brochures to promote AI literacy. However, the present findings serve as a preliminary caution that such interventions may not have the desired effect. Rather than promoting appropriate reliance, we might invite excessive skepticism towards AI's outputs. We suggest that more interactive and thorough interventions might be needed to durably and meaningfully change the way students interact with these tools.

## Acknowledgments

We are grateful to Kristjan-Julius Laak and Madis Vasser for their helpful comments. This work was supported by the Estonian Research Council grants PSG728 and Tem-TA 120 and the Estonian Centre of Excellence in Artificial Intelligence (EXAI), funded by the Estonian Ministry of Education and Research.

# Appendix A. Tasks and ChatGPT recommendations

The tasks and ChatGPT recommendations were presented to the participants in Estonian. The following are translations from Estonian into English. For puzzles with incorrect ChatGPT recommendations the correct answers are marked in bold.

## Correct 1

### Puzzle

There's three rooms, one of which contains a hidden lion. The door of the first room has a sign that says, "The lion is here." The door of the second room has a sign that says, "The lion is not here." The door of the third room has a sign that says, "$3^2 – 2^3 = 2^3 - 3^2$." Only one of the signs is true. In which room is the lion?

### ChatGPT recommendation

Let's analyze the statements on each room's sign:

1. **First room**: "The lion is here."
   - If this is true, then the second sign ("The lion is not here") must be false, creating a contradiction. **It cannot be true.**
2. **Second room**: "The lion is not here."
   - If this is true, then the lion is neither in this room nor in the first room. The lion must be in the third room.
   - The sign on the third room, $3^2 - 2^3 = 2^3 - 3^2$, is not correct ($1 \neq -1$), which means that sign is false. **Only this sign can be true.**
3. **Third room**: $3^2 - 2^3 = 2^3 - 3^2$.
   - The equation is not true. **It cannot be correct.**

**Conclusion**: The second room's sign is true, so the lion is in the third room.

**Answer options**: in the first room, in the second room, the lion can be in any room, the lion can be in either the first or the second room

## Correct 2

### Puzzle

Mikk is buying a sports club membership for the next month, allowing him to train on two days per week. He must immediately choose the two weekdays on which he will attend training. How many different ways can he choose these two days, given that he does not want to train on two consecutive days?

### ChatGPT recommendation

First, we calculate all possible combinations of two days in a week:

$$\binom{7}{2} = 21$$

The pairs of consecutive days are:
(Monday, Tuesday), (Tuesday, Wednesday), (Wednesday, Thursday), (Thursday, Friday), (Friday, Saturday), (Saturday, Sunday), (Sunday, Monday) — a total of 7 pairs.

We subtract the consecutive pairs from all combinations:

$$21 - 7 = 14$$

Thus, the number of possible choices is:

**14**

**Answer options**: 16, 12, 10, 8

**Correct 3**

## Puzzle

Among the boys in the class, no two were born on the same day of the week, and among the girls in the class, no two were born in the same calendar month. If a new student were to join the class, one of these conditions would no longer be met. How many students are currently in the class?

## ChatGPT recommendation

Let there be **b** boys and **g** girls in the class. The following two conditions are given:

1. No boys in the class were born on the same day of the week.
2. No girls in the class were born in the same calendar month.

Since no boys can be born on the same day of the week, the maximum number of boys is 7 (as there are 7 days in a week). Since no girls can be born in the same calendar month, the maximum number of girls is 12 (as there are 12 months in a year).

If a new student were to join the class, one of these conditions would no longer be met. This means that the current number of boys and girls in the class is at its maximum under these conditions.

The maximum number of boys is 7 (since more boys would mean that at least two were born on the same day of the week), and the maximum number of girls is 12 (since more girls would mean that at least two were born in the same calendar month).

**Thus, there are currently 19 students in the class.**

**Answer options:** 18, 20, 24, 25

**Correct 4**

## Puzzle

The teacher asked five of their students how many of them had studied at home the previous day. Pelle said that no one had studied, Elle said that only one person had studied, Mikk said that two had studied, Riho said that three had studied, and Karl said that four had studied. The teacher knew that the students who had not actually studied at home were lying, while those who had studied were telling the truth. How many of these five students had actually studied at home the previous day?

## ChatGPT recommendation

Let **x** be the number of students who studied at home. Pelle said that 0 studied, Elle said that 1 studied, Mikk said that 2 studied, Riho said that 3 studied, and Karl said that 4 studied.

If **x = 0**, Pelle is telling the truth (1 truth, 4 lying). This doesn't work because everyone else would have to be lying. If **x = 1**, Elle is telling the truth (1 truth, 4 lying), which works because one is telling the truth and the rest are lying. If **x = 2**, Mikk is telling the truth (2 truth, 3 lying), but this doesn't work since there should be 2 truthful and 3 lying. If **x = 3**, Riho is telling the truth (3 truth, 2 lying), which doesn't work as there should be 3 truthful and 2 lying. If **x = 4**, Karl is telling the truth (4 truth, 1 lying), which also doesn't work as there should be 4 truthful and 1 lying.

The only possible solution is **x = 1**, meaning that **1 student** studied at home the previous day.

**Answer options**: 0, 2, 3, 4

**Incorrect 1**

## Puzzle

Anni, Pille, Kati, Tiina, and Eerika attended a meeting. Each of them greeted only their acquaintances with a single hug at the beginning. In total, Anni hugged once, Pille hugged twice, Kati hugged three times, and Tiina hugged four times. How many times did Eerika hug? (Note: Each hug is mutual).

## ChatGPT recommendation

Let **A, P, K, T,** and **E** represent Anni, Pille, Kati, Tiina, and Eerika, respectively. Each person hugs only their acquaintances, meaning every hug is mutual.

We know that the number of hugs each person gave is as follows: **A = 1, P = 2, K = 3, T = 4.**

Now, we calculate the total number of hugs. If **x** is the number of hugs Eerika gave, then:

$$1 + 2 + 3 + 4 + x = 2 \times \text{total mutual hugs}$$

Total number of hugs:

$$1 + 2 + 3 + 4 + x = 10 + x$$

Since every hug is mutual, we divide the total number of hugs by two:

$$\frac{10 + x}{2}$$

Since each person could only hug their acquaintances, **x** must be between 0 and 4. The only valid solution is **x = 0**.

Thus, **Eerika hugged 0 times.**

**Answer options:** 5, 4, 3, **2**

## Incorrect 2

## Puzzle

Emma, Ivi, Rita, and Siiri have ages of 3, 8, 12, and 14 years in some order. Both the sum of Siiri and Emma's ages and the sum of Siiri and Rita's ages are divisible by 5. Find the sum of Siiri and Ivi's ages.

## ChatGPT recommendation

The given ages of Emma, Ivi, Rita, and Siiri are 3, 8, 12, and 14 years. We have two conditions:

1. The sum of Siiri and Emma's ages is divisible by 5.
2. The sum of Siiri and Rita's ages is divisible by 5.

Let's check all possible ages for Siiri:

1. **Siiri = 3:** 3 + X and 3 + Y do not give results divisible by 5.
2. **Siiri = 8:** 8 + X and 8 + Y do not give results divisible by 5.
3. **Siiri = 12:** 12 + X and 12 + Y do not give results divisible by 5.
4. **Siiri = 14:**
   - 14 + X and 14 + Y are divisible by 5 only if **X = 12 and Y = 6.**
   - Thus, **Emma = 12 and Rita = 8.**

Siiri and Ivi's ages are **14 and 3**. Their sum is **17**.

**Answer:** 17.

**Answer options**: **26**, 22, 20, 15

## Incorrect 3

## Puzzle

Tiit is convinced that his clock is 5 minutes fast, but in reality, his clock is 10 minutes slow. Leo is convinced that his clock is 10 minutes slow, but in reality, his clock is 5 minutes fast. At one moment, both of them look at their clocks, and Tiit believes that the correct time is 12:00. What does Leo think the correct time is at that moment?

## ChatGPT recommendation

Tiit believes that his clock is 5 minutes fast, but in reality, his clock is 10 minutes slow. So, if Tiit thinks the correct time is 12:00, then according to his clock, the time is:

$$12:00 - 5 \text{ minutes} = 11:55$$

Since his clock is actually 10 minutes slow, the real time is:

$$11:55 + 10 \text{ minutes} = 12:05$$

Now let's check Leo's clock. Leo believes that his clock is 10 minutes slow, but in reality, his clock is 5 minutes fast. When Leo looks at his clock, the time he sees is:

$$\text{Actual time} - 10 \text{ minutes} = 12:05 - 10 \text{ minutes} = 11:55$$

Since Leo's clock is actually 5 minutes fast, the time on his clock is:

$$11:55 - 5 \text{ minutes} = 11:50$$

Thus, if Tiit thinks the correct time is 12:00, then Leo believes the correct time at that moment is **11:50**.

**Answer options**: 11:30, 11:45, 12:00, **12:30**

## Incorrect 4

### Puzzle

An old-fashioned scale is not working correctly. If the weighted item is lighter than 1000 g, the scale shows the correct weight; otherwise, it displays a random number greater than 1000. Five objects, A, B, C, D, and E, were all lighter than 1000 g. The objects were weighed in pairs, and the scale readings were recorded as follows: B + D = 1200, C + E = 2100, B + E = 800, B + C = 900, and A + E = 700. Which of these objects was the heaviest?

### ChatGPT recommendation

We have five objects: A, B, C, D, and E, all weighing less than 1000 g. When weighed in pairs, the recorded results were:
B + D = 1200, C + E = 2100, B + E = 800, B + C = 900, A + E = 700.

The valid equations are: B + E = 800, B + C = 900, A + E = 700. Since B + E = 800 and B + C = 900, it follows that C = 100.
Substituting into A + E = 700: A + (800 - B) = 700, we get A = -100 + B.

Finally, we solve: (900 - B) + (800 - B) = 2100, which results in B = -200.

Thus, **E is the heaviest object**.

**Answer options**: A, B, C, **D**

## Appendix B. Texts presented to the participants

Here we present both the control and intervention text. Horizontal lines indicate where the control text ended. We provide citations, but these were not visible to the participants.

*Before you start solving math puzzles with ChatGPT, please read the following text, which aims to briefly introduce ChatGPT. Please read the text carefully and thoughtfully, even if you have previously encountered and used ChatGPT. Once you have read the text, click the "Next" button in the bottom right corner of the page.*

**A short introduction to ChatGPT**

ChatGPT is a chatbot that was first introduced in 2022 and has since gained rapid popularity. It has been widely adopted in various fields, such as education, economics, customer service, and science, where its use has significantly simplified many tasks.

ChatGPT operates based on user input, which can include questions, tasks, or mathematical operations. It analyzes the user's input and generates an output presented in an understandable language. With ChatGPT, it is possible to solve various tasks, from simple questions to extensive projects. For example, ChatGPT can help generate original text, translate, analyze data, write code, and summarize large amounts of material.

Here are some reasons why ChatGPT is a useful tool for solving language-related tasks (Hariri, 2024):

- **Understanding language.** ChatGPT generates outputs that take into account the context and meaning of words effectively.
- **Personalized responses.** ChatGPT can adapt its responses based on the user's style and previous inputs. This feature enhances the user experience by aligning outputs better with the user's needs.
- **Availability.** As long as there is an internet connection, ChatGPT is always accessible. Its accessibility also extends to its ability to understand and generate responses in multiple languages, including Estonian.
- **Automation.** ChatGPT helps simplify and automate routine tasks, saving the user time and effort.

---

**These lines mark the end of the control text. From here on is the text for the intervention.**

---

**ChatGPT working mechanism**

Each time ChatGPT generates an output, it solves a very simple task: predicting the next word. For example, if I write the sentence "The dog is ___" and ask you which word fits best in the blank, you would likely suggest words such as "an animal," "furry," or "loyal." On the other hand, words like "bottle," "blue," or "fish" would be less likely to fit.

Just as you intuitively know which words are more probable in a given context, ChatGPT also determines this based on its prior training. It has been trained using almost all publicly available text from the internet, allowing it to learn how words are related and which words are most likely to follow others.

However, it's important to note that ChatGPT does not think or experience the world like humans do. At its core, it is simply an extremely large statistical model that responds to user input by calculating the probabilities of the next words.

**Writing prompts**

As previously described, ChatGPT generates output based on user input. This input can consist of four main components (DAIR.AI, 2023):

- **Instructions.** This part explains what task you want ChatGPT to perform. Some examples include: "solve, calculate, write, summarize, evaluate, explain, recommend."
- **Context.** Context provides additional external information that helps ChatGPT better understand the task or the content of the text. This allows it to generate more relevant and accurate outputs.
- **Text.** This is the core content of the input, such as a specific description of an essay you want ChatGPT to write or a passage of text that you want it to summarize.
- **Output preferences.** This allows you to specify the format or style of the response. For example, you can set the word count, paragraph length, or request an output in bullet points.

User input does not have to include all four components. For instance, if you type "If a = b and b = c, then…", ChatGPT will understand that the instruction is to "solve" without needing explicit clarification. However, if you want ChatGPT to generate similar logic problems instead, you need to specify your instructions more clearly.

For any given input, it is useful to consider these components and assess what information ChatGPT might need to produce the most appropriate output.

**Frequent errors in prompts** (Giray, 2023)

- **Unclear input.** Sometimes, user input is not specific enough, resulting in a response that is too broad or general. For example, the prompt "Describe how artificial intelligence has changed the world" is too vague if we specifically want to know about the risks and challenges ChatGPT has caused in the education system.
- **Complex and conflicting instructions.** If instructions are too complicated or contradictory, ChatGPT may struggle to interpret them correctly. An example of conflicting input would be: "Write an essay on technological developments in the 20th century that is detailed enough to be interesting for a computer scientist but also simple enough for a five-year-old to understand."
- **Biased input.** User input can be biased in various ways. For instance, if we want ChatGPT to evaluate the style of a written text, asking, "Is the following text well-written in terms of style?" may nudge ChatGPT towards answering "Yes." A better approach would be to ask neutrally: "Evaluate the style of the following text and provide suggestions for improvement."

**ChatGPT limitations and risks**

Although properly formulating input is crucial for obtaining the best possible response, there are some tasks where ChatGPT may still provide incorrect answers, even if the user input is well-structured. Below is a summary of key limitations of ChatGPT, along with examples.

- **Drawing conclusions about the real world.** ChatGPT does not have direct contact with the real world. Instead, it relies solely on the texts it has been trained on to make inferences about reality. This sometimes leads to mistakes when solving tasks that require reasoning about real-world objects and their properties.

If I have 5 apples right now, but last week I ate 2 apples, then how many do I have left?

If you originally had 5 apples and last week you ate 2 apples, then you have left:

**5 - 2 = 3 apples.**

**Example 1** (task inspired by Diaz, 2024). In the above example, ChatGPT misunderstands the concept of time (now, last week) and mistakenly assumes that it is a simple arithmetic problem.

- **Mathematical and logical reasoning.** ChatGPT may sometimes incorrectly solve even simple problems that require mathematical and logical reasoning.

How many words and how many numbers are in the following list: cat, 10, dog, 60, cow, 40, squirrel, 70, spruce, 15, book, 32.

The list contains 7 words and 5 numbers.

**Example 2**.

The five given cards must be arranged in such a way that the resulting 9-digit number is the smallest possible. Which of these cards must be the rightmost in this case? The cards are: 4, 8, 31, 59, 107.

To form the smallest possible 9-digit number, the cards must be placed in ascending order, considering the value of each number:

- The smallest number is 4, followed by 8, then 31, then 59, and finally, the largest number, 107.

If we arrange these to create the smallest number, it is logical that the largest card, 107, must be the rightmost, as placing it at the last position increases the overall value the least.

**Answer: The rightmost card must be 107.**

**Example 3** (task from Teaduskool, 2022). 107 should be the leftmost number, and 8 should be the rightmost number.

- **Hallucination.** ChatGPT sometimes produces completely fabricated information – a phenomenon known as hallucination. This can include generating false references and historical facts. Additionally, it may struggle to distinguish between fiction and reality.

Who was Estonia's first Olympic silver medalist?

Estonia's first Olympic silver medalist was **Jaan Tõnisson**, who won the silver medal at the 1920 Summer Olympics in Antwerp, Belgium. Tõnisson competed in Greco-Roman wrestling in the heavyweight category.

**Example 4**. Estonia's first Olympic silver medalist was Martin Klein in 1912, and Jaan Tõnisson was not a wrestler but an important statesman.

ChatGPT can hallucinate in response to various types of questions. However, it is especially easy to trigger this when user instructions require a very narrow and specific output.

Name five cities in Harju County that start and end with the same letter.

There are five cities in Harju County that start and end with the same letter:

1. Keila
2. Saue
3. Maardu
4. Paldiski
5. Jõelähtme

**Example 5.**

Answer with one word: In which sport did Napoleon win a gold medal at the 1980 Olympic Games?

Football.

**Example 6.**

- **Discriminatory outputs.** ChatGPT's outputs depend on the texts it was trained on. These texts may contain various social biases related to people's social status, gender, race, and age, and during training, the AI learns to incorporate these biases into its outputs. As a result, ChatGPT's responses may be politically biased or include gender stereotypes that exist in society.

The CEO of a large company went to the restroom before going to the manicurist. The men's and women's restrooms were separate. Which restroom did they go to?

Based on the sentence, we can infer that the CEO is a man because, in such contexts, when gender is not explicitly specified and the general term "CEO" is used, it is often assumed to refer to a man. Therefore, they went to the men's restroom.

If this sentence is part of a story and you want to provide a clearer answer, it could depend on additional context or the explicitly stated gender of the character. In this case, the assumption would be that they used the men's restroom.

**Example 7.**

In addition to knowing that ChatGPT can make mistakes, it is important to understand that some of its characteristics make detecting these errors difficult.

- **Excessive confidence.** ChatGPT presents outputs with high confidence, even when they contain misinformation. Instead of saying, "I'm not sure," it delivers responses in a very convincing and fluent manner, making it harder to notice errors in its outputs.
- **Errors are unexpected.** ChatGPT's mistakes can appear unpredictably, which often makes them go unnoticed. It may clearly and accurately explain complex quantum physics concepts one moment and then make mistakes in the simplest logic problems the next. This inconsistency in response quality poses a risk when using ChatGPT. Based on previous correct answers, we might assume that ChatGPT can correctly solve subsequent tasks—even when those responses actually contain errors.

**Tips for using ChatGPT**

- **Ask multiple times.** ChatGPT's responses are based on probability calculations, meaning it may provide different answers to the same question. Some of these answers might be unlikely or incorrect. Therefore, it is helpful to ask the same question multiple times with slightly different wording so you can choose the most suitable response.
- **Ask for the reasoning.** When solving problems that require logical thinking, include an instruction in your input asking ChatGPT to explain its reasoning (e.g., by adding "Solve step by step" at the end of your prompt). This makes it easier to understand how ChatGPT arrived at its answer and helps identify if the reasoning is flawed.
- **Verify facts.** If you're unsure whether the information in ChatGPT's response is accurate, always verify it using other sources. One option is to search on Google. Another is to ask ChatGPT to provide references in its output so you can check their reliability yourself.

**Remember that ChatGPT is a tool, not a replacement for thinking!**

Appendix C. Participants' reliance across puzzles

| Puzzle | Over-reliance | Under-reliance | Human error | Correct solutions |
|---|---|---|---|---|
| Correct 1 | – | 16 (44.4%) | – | 20 (55.6%) |
| Correct 2 | – | 15 (41.7%) | – | 21 (58.3%) |
| Correct 3 | – | 5 (13.9%) | – | 31 (86.1%) |
| Correct 4 | – | 5 (13.9%) | – | 31 (86.1%) |
| Incorrect 1 | 17 (47.2%) | – | 10 (27.8%) | 9 (25%) |
| Incorrect 2 | 21 (58.3%) | – | 5 (13.9%) | 10 (27.8%) |
| Incorrect 3 | 18 (50%) | – | 17 (47.2%) | 1 (2.8%) |
| Incorrect 4 | 19 (52.8%) | – | 15 (41.7%) | 2 (5.6%) |
| *M* | 18.8 (52.1%) | 10.3 (28.5%) | 11.8 (32.6%) | 15.6 (43.3%) |

*Note.* Over-reliance = participant used the incorrect recommendation; Under-reliance = participant did not use the correct recommendation; Human error = participant ignored the incorrect recommendation but subsequently answered incorrectly.